\documentclass{mem}
\usepackage{natbib}\usepackage{txfonts}\usepackage{balance}
\usepackage{graphicx}
\usepackage[a4paper,breaklinks,dvipdfm]{hyperref}
\idline{75}{282}
\begin{document}
\def\teff{$T\rm_{eff }$}
\def\kms{$\mathrm {km s}^{-1}$}

\title{
Galactic chemical evolution revisited
}

   \subtitle{}

\author{
M. \,Haywood\inst{1} 
          }

  \offprints{M. Haywood}

\institute{GEPI, Observatoire de Paris, GEPI, CNRS, Universit\'e Paris Diderot, 5 Place Jules Janssen, 92190 Meudon, France
\email{misha.haywood@obspm.fr}
}

\authorrunning{Haywood}

\titlerunning{Galactic Chemical evolution revisited}

\abstract{
Standard chemical evolution models based on long-term infall are affected by a number of problems, evidenced
by the analysis of the most recent data.
Among these: (1) models rely on the local metallicity distribution, assuming its shape is 
valid for the entire Galaxy, which it is not; (2) they assume that the solar vicinity abundance patterns 
resulted from a unique chemical evolution, 
which it does not; (3) they assume the disk is a single structure with chemical properties that are a smooth function of
the distance to the galactic centre, which it is not.
Moreover, new results point to a thick disk being as massive as the thin
disk, leading to a change of paradigm in the way we see the formation of these structures.
I discuss these various issues, and, commenting on Snaith et al. (2014), how a closed box model offers 
an interesting approximation to the galactic 
chemical evolution, by providing the conditions in which 
large amounts of gas are available in the disk at high redshift.
The novel way presented in Snaith et al. (2014) to derive SFH from stellar abundances is also discussed, providing a measurement
of the SFH of old populations that is valid for the entire Galaxy. The derived SFH shows
that the formation of the thick disk has been the dominant epoch of star formation in our Galaxy.
}
\maketitle{}

\section{Introduction}

In the last three decades, the requirement that chemical evolution models of the Galaxy 
fit the local metallicity distribution and the radial metallicity gradient has dictated a scenario where the disk 
is built inside-out through slow accretion of gas. Models have now reached fair sophistication
and claim to be able to reproduce local chemical 
patterns through one infall models (e.g, \cite{naa06}), and up to 
as many infall episodes as there are populations in the solar vicinity \citep{mic13}.
In pratice, models differ very little from each other, the common thread being 
a limited production of stars early in the life of the Galaxy, in order to resolve the so-called
G-dwarf problem \citep{van62}. As a consequence, the production of stars in the thick disk phase is, by construction, limited to 15-20\%, at most 30\%.
Following the same line of thought, Fraternali \& Tomassetti (2013) derived the accretion history of the MW, 
and contended that the MW produced two thirds of its stars after z=1.
In the following pages, I review an emerging new context where recent observational results and new 
modelling of the chemical evolution of the Milky Way point to a radically different picture.

In the next section, I present a series of problems affecting standard models.
In section 3, I present the new observational scene, and discuss evidences showing that a closed box model, provides, to zero-order 
approximation, a good description of the solar vicinity data, as recently presented in Snaith et al. (2014). In section 4, I discuss a number of issues directly related to chemical evolution
models, such as radial mixing and the inside-out scenario. I conclude in section 5.

\section{Standard chemical evolution of our Galaxy}

\subsection{Standard chemical evolution}

The  overwhelming majority  of present  chemical evolution  models are based  on the  observation that  the disk  in the  
solar neighbourhood contains far too few metal-poor stars -- here, 'metal-poor' means about 1/3 of the solar 
metallicity -- to have formed from  an intense initial phase of low metallicity gas consumption.  This is the  
so-called 'G-dwarf' problem.   To overcome this problem, most studies have  endorsed the idea that  the gas must  have been 
conveyed onto the disk on a  relatively long time scale, maintaining a small gas reservoir, a limited dilution and, 
consequently, a rapid rise of the metallicity in the ISM before many stars formed.  
The infall paradigm constitutes the basis of current Galactic  Chemical  Evolution  codes  \citep[e.g.][]{chi97,naa06,mic13}. 

The density  of gas  is then  used to  calculate the  star  formation rate usually assuming a Schmidt-Kennicutt law. 
The form of the gas infall law is then ajusted in order to fit the local  metallicity distribution function (hereafter MDF), and the chemical  
evolution tracks generated are finally confronted against metallicity-[$\alpha$/Fe] distributions. 
It is the fit to the local MDF of dwarfs that determines the infall time scale:
the longer the time scale, the fewer the number of low metallicity dwarfs, and the narrower the MDF. 
However, the time scale measured on the local MDF is meaningful as long as the MDF is invariant over the whole disk, 
a condition we shall see is most probably not verified.

\subsection{Problems and shortcomings}

We list here some fundamental limitations and problems standard chemical evolution models of the Milky Way are likely to meet 
in the new observational context that has emerged in recent years.

(1) {\it Models assume that the solar vicinity is representative of the entire disk, which it is not}.
Constraints available at the solar vicinity are not valid over the entire disk. 
The local MDF is made from a mixture of thick and thin disks stars, in the relative proportion of 5-10\%.
Recent  measurements show that the thick disk scale length is significantly 
shorter than that of the thin disk (see Bensby et al. 2011, Cheng et al. 2012, Bovy et el. 2012a), 
with respectively $\sim$2 kpc and $\sim$3.6 kpc, so that we expect this mixture to vary with galactocentric distance,
and the MDF to vary accordingly.
Present models, when fitted to the solar vicinity, are thus not generalizable and cannot be utilised to describe the 
chemical evolution of the disk. 
We caution that, if the difference in scale lengths is confirmed, the local metallicity distribution {\it should not} be used as it has been 
in the last decades to set constraints on galactic chemical evolution models.
In fact, the majority of low-mass, intermediate metallicities, missing G-dwarf stars is expected to reside in the 
inner disk, which means they are not missing, but only under-represented in the solar vicinity.
Hence, models that aim to describe the whole MW, as  standard models do, cannot be calibrated on the 
solar vicinity ratio of metal-poor to metal-rich stars.

(2) {\it Models assume that the solar vicinity chemical patterns are the result of a single chemical evolution, which they
are not}. Figure 1 illustrates the two sequences that have been attributed to the thin and thick disks. 
The age scale shows that stars with similar ages (at [$\alpha$/Fe]$\sim$ 0.1 dex) in the thin and thick disk sequences have metallicities that differ by about 0.5 dex (respectively [Fe/H]$\sim$-0.6 dex and $\sim$-0.1 dex), 
implying that they cannot have resulted from the same chemical evolution history (see Haywood et al. 2013).
Moreover, low metallicity stars in the thin disk have been shown to possess mean orbital radii and kinematics compatible
with an origin in the outer disk (Haywood, 2008), while the more extended SEGUE sample confirms a mean 
orbital radii greater than $\sim$9kpc (Bovy et al. 2012a, Fig. 7), pointing also to an outer disk origin.

(3) {\it Models assume that the disk is a single structure whose chemical properties change smoothly with radial 
distance from the galactic centre, which it is not}. This assumption has, in particular, been used to
justify radial mixing by churning \citep{sch09}, see section 4. However, observations in the recent years have shown that 
the Sun is in a transition region that separate two disks with different properties. Hence, the metallicity 
gradient is steep in the solar vicinity, but is much flatter both in the inner (R$<$7kpc) and outer (R$>$9-10kpc) disks. 
The disk is structured, and is not a single system with properties smoothly varying with radial distance to
the Galactic centre. There is actually more chemical continuity between the thick and thin disks than between 
the inner (as defined by the sum of the thick and thin disks within 10~kpc from the Galactic centre) and outer disks, see Haywood et al. (2013). Similar features have been observed on other disk galaxies, with the best example provided by M33 (see Bresolin et al. 2012).

(4) {\it Chemical patterns in  [$\alpha$/Fe]-[Fe/H] plane are a degenerate function of the star formation history.}
In other words, a range of star formation histories is compatible with even the most accurate spectroscopic measurements
available (see Snaith et al. 2014, in preparation). 
Therefore, fitting the observed chemical trends does not guarantee that the assumed SFH is correct. Since 
in standard chemical evolution models, the accretion is directly linked to the SFR through the Kennicutt 
law, this implies that the deduced accretion history has no reason to be correct either.

\begin{figure*}
\includegraphics[width=7.cm,angle=90]{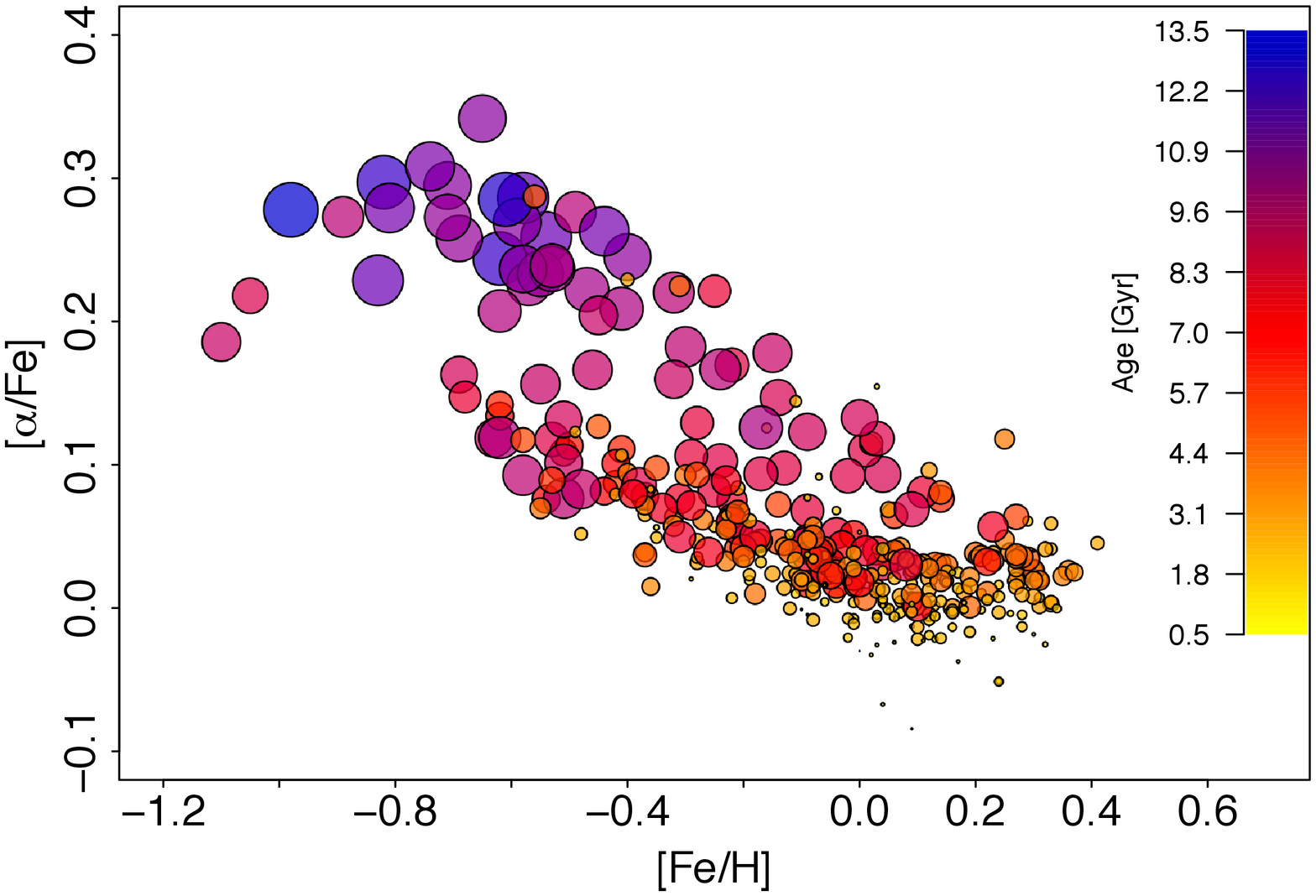}
\caption{ [$\alpha$/Fe] $vs$ [Fe/H] for the stars in the sample of
Adibekyan et al. for which a robust age could be derived. The color and
the size of the symbols both code the age of the stars, to emphasize
the age stratification of the distribution of stars within this plane. From Haywood et al. (2013) }
\label{alphafehage}
\end{figure*}

\section{Revisiting galactic chemical evolution}

\subsection{The new observational scene}

Recent observational results have changed significantly our vision of stellar populations as sampled
in the solar vicinity, but also on a larger scale. The disk cannot be described as a unique structure 
with properties smoothly varying with distance from the galactic centre. 

{\it (1) Structural aspects}

There are now a number of converging studies that show the thick disk has a scale length 
significantly smaller than the thin disk scale length (Bensby et al. 2011, Cheng et al. 2012, Bovy et al. 2012a). 
Bensby et al. (2011) proposed that the scale length of the thick disk is about 2kpc, a result that was 
supported by the analysis of Cheng et al. (2012) using SEGUE data, and Bovy et al. (2012a). 
The common characteristics of these studies is that they use $\alpha$ elements to discriminate thin 
from thick disk stars, which is indispensable to lift off the degeneracy that otherwise affect any
attempt to estimate the relative contributions of these two populations from star counts only. 
The MDF being a composite mixture of thin and thick disk stars, the immediate consequence is that 
the relative proportion of the two populations is expected to vary as a function of distance to the
galactic centre. 
This also seems to be confirmed by the first APOGEE results, which show that stars selected to have median orbital radii
between 4 and 7 kpc are roughly in equal proportions below  [Fe/H]$\sim$-0.2 dex  (mainly thick disk with  [$\alpha$/Fe]$>$0.15 dex) 
and above this limit 
(mainly thin disk with [$\alpha$/Fe]$<$0.15 dex).
The shape of the histograms in Fig. 14 of \cite{and13} is the illustration that the mean 
metallicity does not vary smoothly with distance to the galactic centre. In fact, the peak metallicity
is shown to decrease from the solar circle in both the directions of the inner disk at [4-7]kpc and the outer disk at [11-13]kpc, which is
hardly understandable in terms of smooth gradients. More likely, the mean metallicity decreases 
towards the inner disk because of the mounting importance of the thick disk, 
and decreases towards the outer disk because of the transition to a disk population with 
different overall characteristics. 
In that respect, the APOGEE results confirm the trends seen in other studies, that 
the mean metallicity of stars beyond R=9-10~kpc is about -0.3dex, i.e decreasing abruptly in 1 or 2kpc from the 
mean solar value (-0.05 dex). The change of population as one moves outwards is also strickingly illustrated in the 
sample of Bensby et al. (2013, Fig. 26), where it is shown that the sample of stars with mean orbital radii R$<$7 kpc contains almost no 
outer thin disk objects of low metallicities, while the stars with R$>$9kpc contain no metal-rich thin disk and no 
thick disk objects. 
It is only at the solar radius that all types of stars are represented, supporting the analysis of Haywood et al. (2013)
that the Sun is in a transition region between the inner disk, composed of the thin and thick disks, which are both 
radially limited to 9-10kpc from the galactic centre, and the outer disk at R$>$10kpc.
Another consequence of the shorter thick disk scale length is that this population is much more massive than 
previously thought. Considering the surface densities estimated from the SEGUE data (Bovy et al. 2012b), scale lengths
of the order of 2 and 3.6kpc induce stellar thick and thin disks of roughly equal masses.

{\it (2) Evolutionary aspects}

The inner disk is composed of the thick and thin disks, which are essentially in 
continuity (Bovy et al. (2012a), Haywood et al. (2013)), although a marked transition 
between the two is encrypted in the change of the evolution of $\alpha$ elements 
and metallicity with age (see Haywood et al. (2013) and Fig. 2(a), for the case of silicon, Snaith et al. 2014), due to a change in the regime 
of star formation at that epoch (see Fig. 2(b), and next subsection). A tight age-metallicity relation has 
been found in the thick disk (Haywood et al. 2013), testifying 
that the ISM at this epoch was well mixed. 
The data show that the metallicity steadily increased
during a period lasting 4 to 5 Gyr, driving the metallicity and $\alpha$ abundances well in
accordance with the thin disk 8 Gyr ago.
This is in line with the observation of thick gaseous disk at high 
redshift with scale height of $\sim$ 1~kpc (Elmegreen et al. 2004, Bournaud et al. 2009)
and disk-like kinematics, which could be the progenitors of present day thick disks (Genzel et al. 2006).
Spectroscopic observations also show high  velocity
dispersions  in their  gas, similar  to those  measured in  the present
stellar thick  disk of the  Milky Way \citep{swi11, leh13}.

According to Haywood et al. (2013), the outer disk started to form stars 10 Gyr ago, i.e when the thick disk formation 
was still on-going in the inner Galaxy (R $<$ 10kpc). The similarity in $\alpha$-element abundance between 
the thick disk and the outer disk at identical ages (10 Gyr) suggests that the gaseous material from which stars 
formed in the outer disk could have been polluted by outflows from the forming thick disk. 
The substantially lower metallicity of the
oldest outer disk stars ([Fe/H]$\sim$ -0.7 dex) (compared to the already high metallicity reached by the thick
disk at this epoch) also suggests that this gas mixed with more pristine gas present in the disk outskirts, in a 
scheme where the main accretion of gas could have gradually migrated from the inner disk (R$<$10kpc) to the
outer regions.

These various properties suggest that we have two different structures whose formation history and chemical 
evolution may be different. Such overall disk structure has been observed in other galaxies. M33 is known
to possess a flat radial metalliciy gradient beyond about 10kpc (Cioni 2009). This is accompanied by a break in the density
profile and an upturn in the mean age of the underlying stellar population (Barker et al. 2011). Similar flattenings of metallicity
distributions are observed in a number of disks (Bresolin et al. 2012).

\begin{figure*}
\includegraphics[trim=100 180 80 280,clip,width=7.cm]{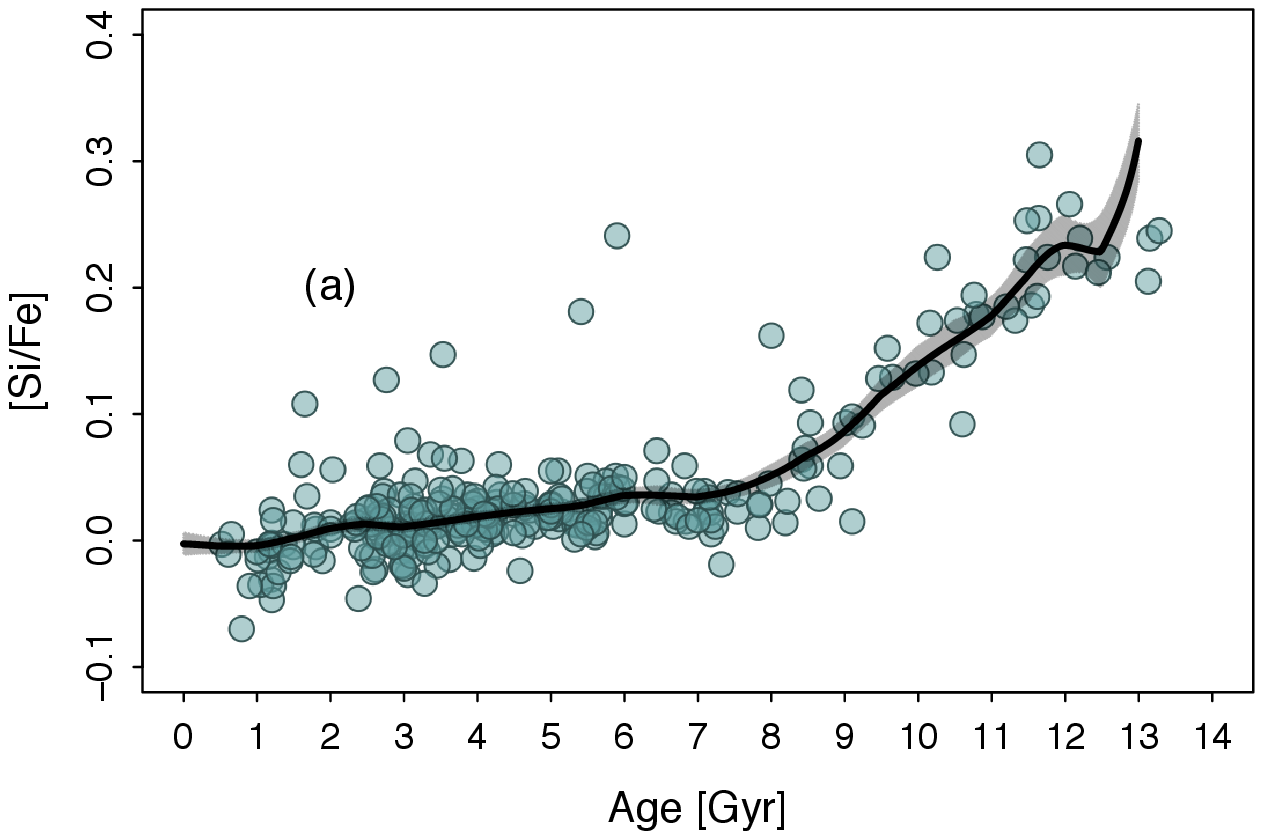}
\includegraphics[trim=100 180 80 280,clip,width=7.cm]{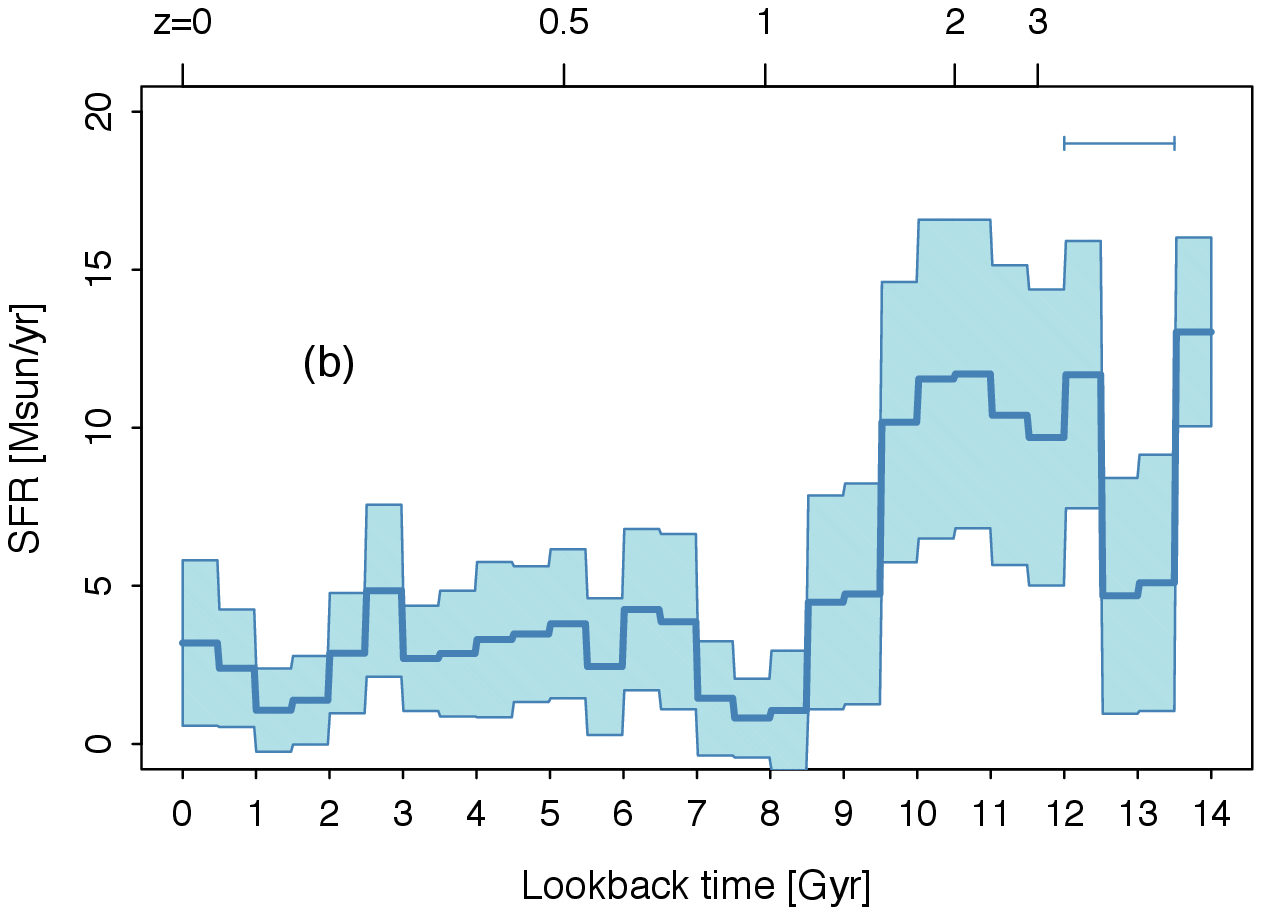}
\includegraphics[trim=100 180 80 300,clip,width=7.cm]{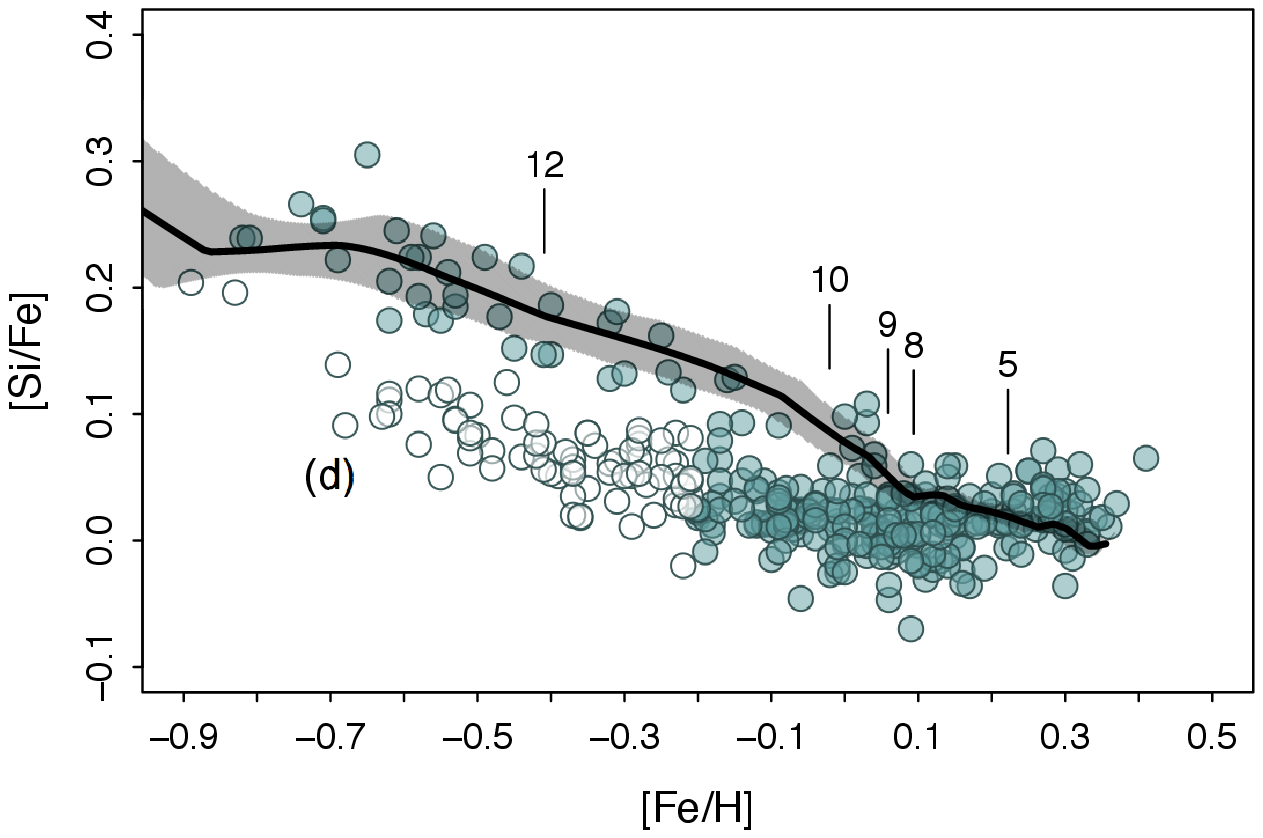}
\includegraphics[trim=100 180 80 300,clip,width=7.cm]{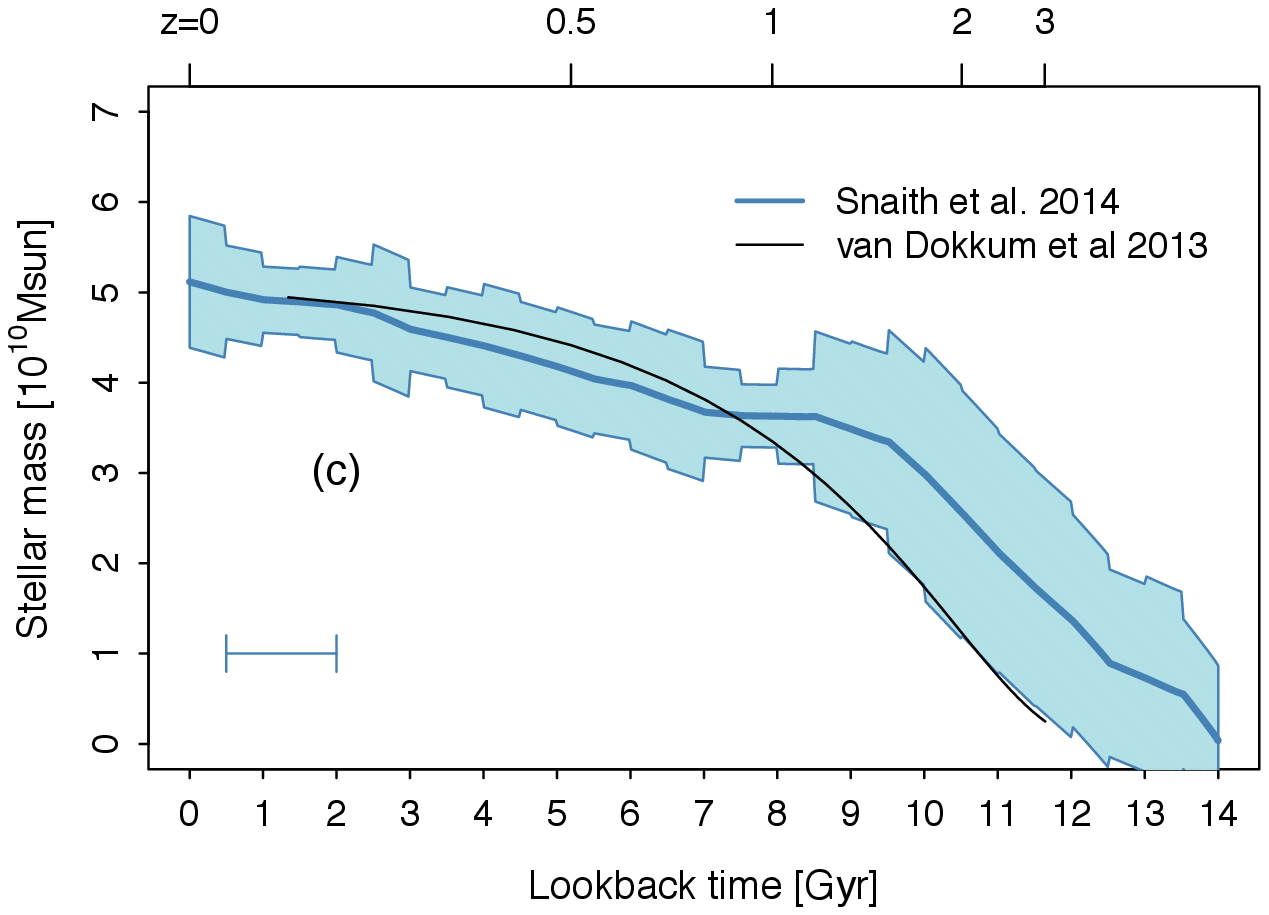}
\caption{Results from Snaith et al. (2014) \emph{Panel~(a): } [Si/Fe] vs age relation (black curve) as determined by fitting a closed-box model to (inner disk) data from Adibekyan et al (2012) and Haywood et al. (2013; green circles). \emph{Panel~(b): } The star formation history deduced from the fit in (a).  \emph{Panel~(c): } Cumulative stellar mass as a function of redshift for the MW from (b) compared to that of Milky Way-analog galaxies from van Dokkum et al. (2013; black curve). 
\emph{Panel~(d): } [Si/Fe]-[Fe/H] chemical evolution track deduced from the model. No fit is made. Tick marks indicate the age of the model. See Snaith et al. (2014) for details.}
\label{alphafehage}
\end{figure*}

\subsection{A ``new'' model: the closed-box}

The finding that the thick disk is as massive as the thin disk suggests that large amounts of gas
were available early in the formation of the disk in order to sustain the production of large 
amounts of intermediate metallicity stars. Recent results 
are showing that the progenitors of the Milky Way already had built half their stellar mass at z$\sim$1.5 \citep{van13}. 
This suggests that it may be interesting to model the MW disk as a system where the SFR is not limited by 
the available amount of gas.
This is also very much in line with recent studies suggesting that the SFR at these epochs may not be 
directly linked to gas accretion, being severely reduced by stellar feedback (Hopkins et al. (2013)). 

 The closed-box model is an idealized model where all the gas is assumed to be contained in the system
from the beginning, and can be 
seen as a zero-order approximation of a scenario where most of the gas is accreted onto the disk very early in the 
life of the Galaxy -- 
which is not meant to say that zero-accretion occured later 
in the evolution of the disk, but simply that we seek how far the main features of the Galactic chemical 
evolution can be described, in first approximation, by a closed-box. 
Interestingly, a closed-box model with a peak metallicity at -0.05 dex (the solar vicinity observed
value) has a median metallicity 
at -0.3 dex, which is close to the transition between the thick and thin disks.

A closed-box model, with ingredients similar to other chemical evolution models,
can be used to fit the solar vicinity chemical trends, see Snaith et al. (2014). 
In fitting solar vicinity data, it is crucial to reproduce the age $vs$ chemical abundances relations, 
because, as already mentioned, models are degenerated in the [$\alpha$/Fe]-[Fe/H] (see Snaith et al., 2014).
Moreover, the chemical evolution of the thin disk is a very limited segment of the ([$\alpha$/Fe],[Fe/H]) parameter
space (see Fig.(2)d), while $\alpha$ elements are a tight function of age over the whole age range (see Fig. 2(a)).

Fig. 2(a) shows the result of fitting a model to the Age $vs$ [Si/Fe] distribution as done in Snaith et al. (2014), 
through iterative search of the correct SFH. The corresponding best fit SFH is given in Fig. 2(b), where the
SFH is normalized such that the integrated SFH corresponds to a total stellar mass of 5.10$^{10}$M$_{\odot}$.
Fig. 2(b) shows that half the disk stellar mass was formed during the thick disk phase. This is in good agreement
with estimates of the thick disk mass from structural parameters.
Interestingly, the result of Fig. 2(b) is very similar to the SFH obtained from Milky Way progenitors at high redshift for which ...
{\it ``the implied star formation rate is approximately 
constant at 10-15 M$_{\odot}$/yr from z$\sim$2.5 to z$\sim$1 then 
decreases rapidly to $<$  2 M$_{\odot}$/yr  at z=0.''}, van Dokkum et al. (2013).
Fig. 2 (d) shows how the model reproduces the thick disk sequence in the ([Si/Fe], [Fe/H]) plane.
The thin disk sequence is less well described, as expected, because it is not 
temporal, but reflects a dispersion due to the position of the Sun at the interface of the inner and outer
disks.

These results are evidence that the disk chemical evolution is well described by a system
where most of the gas is acquired very early, while the thin disk star formation rate can
be maintained by the gas leftover from the thick disk phase and the gas recycled from
previous generations. 
In this context, is it worth quoting Hopkins et al. (2013), saying that the feedback {``\it produces a reservoir of gas 
that leads to flatter or rising late-time star formation
histories significantly different from the halo accretion history''}.

\section{Other issues}

\subsection{No radial mixing at the Sun}

Much attention has been focused in the recent years on the possible importance of radial
mixing in the sense defined by \cite{sell02}.
The present (lack of) evidences have been discussed in \cite{hay13}, and its 
worth summarizing them here. 
Claims in favor of radial mixing are based on the argument that, assuming a radial gradient
of about -0.07dex/kpc, stars that are $\sim$0.3 dex more metal-rich or metal-poor than the mean 
solar vicinity metallicity (about -0.05 dex) must have come from farther distances than 4kpc. 
If  angular momentum conservation is assumed, this would imply rotational velocities for migrating stars 
that are not seen on solar vicinity stars. Therefore, it has been proposed \citep{sch09} that some mechanism as the one proposed by 
\cite{sell02} must be at work. 
There are strong doubts that this reasonning is verified, simply because stars that are 
0.3 dex more metal-poor or metal-rich than the solar vicinity are  massively present at
just 2~kpc from the solar orbit. If radial mixing was effective across the solar 
orbit, we would expect to see these stars in important number, while they represent only 
a few percents at the solar vicinity.  
The most recent data show that indeed, the local ($<$2kpc) gradient is steep, 
with the mean metallicity at R$\sim$10kpc being $\sim$-0.3 dex, while the APOGEE data
show that the mean metallicity of the {\it thin} disk at R$\sim$7kpc is $\sim$0.2 dex, 
which implies in any case that the gradient is steeper than $\sim$ -0.12 dex/kpc.
Note that a strong gradient is in itself evidence against significant mixing. 
The U dispersion measured on metal-poor thin disk stars is of the order of 50km/s, which is sufficient to ensure 
radial excursion of the order of 2kpc, putting the solar vicinity within reach of inner or outer disk 
stars of required metallicity during their radial epicycle oscillation. 
Hence, as argued in Haywood et al. (2013), simple effects of blurring (radial 
excursions of stars on their orbits due to epicycle oscillations) are sufficient to contaminate the 
solar vicinity with a few percent of stars coming from either the inner or outer disks. 
That is, we suggest that the spread in metallicity observed at the solar vicinity is due 
essentially to the Sun being at the frontier of two systems with markedly different mean 
metallicities.
This is even more obvious when chemical trends of stars in the solar vicinity are inspected as a function 
of mean orbital radius, where it is seen that for R$_m <$ 7 kpc, metal-poor thin disk objects become
rare, while at R$ >$ 9 kpc, thick disk stars are nearly absent.
Only stars that have mean orbital radii between 7 and 9 kpc show both characteristics, confirming that 
the Sun is at the interface between the two systems (inner and outer disks).

\subsection{Inside-out}

The inside-out paradigm has been proposed first by Larson (1976), and then adopted as an adhoc
prescription in chemical evolution models in order to ensure the presence of gradients, by adopting some radial dependence of the infall law. 
Since the radial metallicity gradients are the main motivation for invoking the inside-out paradigm in the Milky Way, it is worth recalling the 
observational evidences. 
The radial distribution of metallicity as skethed in section 3.1 is better described by the combination of disks with 
different mean metallicities and radial extent but no gradient, than a single disk with a smooth gradient.
No gradient has been found within the thick disk, which is shown to have a radial extent of about 10~kpc
(Bensby et al. (2011), Cheng et al. (2012)).
The thin disk within 6-7kpc from the galactic centre is well described by a mean metallicity of about +0.2 dex
in the APOGEE data. The thin disk beyond 9-10kpc seems to have a mean metallicity of -0.3 dex, 
with a negligible, or small, gradient. 
As already mentioned, the steep gradient seen in the solar region can be interpreted as resulting from the transition 
between two systems, the metal-rich and the metal-poor thin disks. 
The analysis of the SEGUE data shows (Bovy et al. 2012a)  that the scale length of the thick disk
(as defined by stars with [$\alpha$/Fe]$>$0.25 dex) does not increase with decreasing alpha abundance, although the duration of the formation of this 
population is rather long ($\sim$ 4 Gyr). Similarly, no systematic increase of the scale 
length of the thin disk is seen within 8 Gyr. 
Overall, there is therefore very little evidence for a systematic (and progressive) increase of the scale length of the disk
with time.

\section{Conclusions}

We summarize the above discussion as follows:
(1) recents observational results suggest that a very significant revision of the chemical evolution
models is necessary, because the importance of the thick disk has been severily underestimated.
We caution in particular that the local MDF {\it should not} be 
used to constrain models. 
(2) The Sun is at the interface of two systems, the inner and outer disks, and the chemical 
patterns observed in the solar vicinity are therefore a mix of patterns generated by two
different chemical evolutions.
(3) New results both for the Milky Way and Milky Way progenitors at high redshifts, suggest that the disk sustained high rates
of star formation that formed the thick disk.
A closed box model, i.e a system 
where large amount of gas is present and can sustain the high rate of star formation at these epochs, 
produces a good match to the local chemical trends.  
(4) New arguments are emerging that support the idea that the SFH is not directly related
to the accretion history, see Hopkins et al. (2013). The derived SFH from Snaith et al. (2014) 
shows that the thick disk formed $\sim$ 50\% of the stellar mass of the Milky Way, in accordance with extragalactic
studies \citep{van13,muz13}, but at variance with the chemical evolution models with infall developed in the last 
decades.

Taking into account the fact that there are now substantial evidences that the Milky Way has a very small classical bulge ($<$10\% of the disk)
\citep{she10,kun12,dim14} or possibly no classical bulge at all -- in this respect, the Milky Way is similar to most of disk galaxies
in the local Universe \citep{kor10, fis10} --  at the epoch when galaxies were reaching the peak of their
star formation, the only significant population to have formed in our Galaxy was the thick disk. 
With star formation intensities $\sim$5 times the present value, and an overall stellar mass of about
50\% of the present total stellar mass, the thick disk has arguably been the dominant epoch of star formation
in the Milky Way.

\begin{acknowledgements}
I am very grateful to the organizers for inviting me to this interesting conference.
I am indebted to Paola Di Matteo, Matt Lehnert and Owain Snaith for insightful discussions 
and enjoyable collaboration in the last year and a half. 
\end{acknowledgements}

\bibliographystyle{aa}

\end{document}